\documentclass[10pt,twocolumn,letterpaper]{article}

\usepackage{cvpr}              

\definecolor{cvprblue}{rgb}{0.21,0.49,0.74}
\usepackage[pagebackref,breaklinks,colorlinks,allcolors=cvprblue]{hyperref}

\def\paperID{15361} 
\def\confName{CVPR}
\def\confYear{2025}

\title{Rethinking Decoder Design: Improving Biomarker Segmentation Using Depth-to-Space Restoration and Residual Linear Attention}

\author{
    Saad Wazir$^{1}$ \quad Daeyoung Kim$^{1}$ \\
    \texttt{saad.wazir@kaist.ac.kr, kimd@kaist.ac.kr} \\
    $^{1}$School of Computing, KAIST, Republic of Korea
}

\begin{document}

\maketitle

\AddToShipoutPicture*{%
  \put(0,50){%
    \makebox[\paperwidth]{\centering\footnotesize
    \parbox{\dimexpr0.9\paperwidth}{\centering
    © 2025 IEEE. This paper has been accepted to the IEEE/CVF Conference on Computer Vision and Pattern Recognition (CVPR) 2025.\\
    Personal use permitted. For other uses, permission must be obtained from IEEE. \href{https://openaccess.thecvf.com/content/CVPR2025/html/Wazir_Rethinking_Decoder_Design_Improving_Biomarker_Segmentation_Using_Depth-to-Space_Restoration_and_CVPR_2025_paper.html}{CVPR 2025 Open Access version} }%
    }%
  }%
}

\begin{abstract}
Segmenting biomarkers in medical images is crucial for various biotech applications. Despite advances, Transformer and CNN based methods often struggle with variations in staining and morphology, limiting feature extraction. In medical image segmentation, where datasets often have limited sample availability, recent state-of-the-art (SOTA) methods achieve higher accuracy by leveraging pre-trained encoders, whereas end-to-end methods tend to underperform. This is due to challenges in effectively transferring rich multiscale features from encoders to decoders, as well as limitations in decoder efficiency. To address these issues, we propose an architecture that captures multi-scale local and global contextual information and a novel decoder design, which effectively integrates features from the encoder, emphasizes important channels and regions, and reconstructs spatial dimensions to enhance segmentation accuracy. Our method, compatible with various encoders, outperforms SOTA methods, as demonstrated by experiments on four datasets and ablation studies. Specifically, our method achieves absolute performance gains of 2.76\% on MoNuSeg, 3.12\% on DSB, 2.87\% on Electron Microscopy, and 4.03\% on TNBC datasets compared to existing SOTA methods. Code: \href{https://github.com/saadwazir/MCADS-Decoder}{https://github.com/saadwazir/MCADS-Decoder}

\end{abstract}

\section{Introduction}
\label{sec:intro}
Segmenting biomarkers in medical images is crucial for biomedical applications involving the evaluation of biomarkers like glands and nuclei for accurate diagnoses. Despite its importance, medical image segmentation faces challenges due to limited samples and the variability in biomarker appearances and sizes. In medical image segmentation, where datasets often have limited sample availability, recent SOTA methods achieve higher accuracy by leveraging pre-trained encoders, whereas end-to-end models tend to underperform. This underperformance is due to challenges in effectively transferring rich multiscale features from encoders to decoders, as well as limitations in decoder efficiency.

Convolutional Neural Networks (CNN) based models like UNet \cite{inp:17} and its advanced variants (nested \cite{a:5}, recurrent \cite{a:25}, dual \cite{a:26, inp:25} encoder-decoder structures) have improved segmentation accuracy. However, these models often struggle to capture long-range dependencies, a limitation mitigated by techniques like dilated convolutions \cite{a:31} and attention mechanisms \cite{a:6, a:36, a:37, a:38}. While CNNs are crucial for vision tasks, their reliance on localized receptive fields limits their ability to capture long-range dependencies in images, leading to only modest improvements and highlighting the need for further accuracy enhancements.

Recent transformer based models, such as Vision Transformers, excel at capturing long-range interactions but lack precision when modeling local contexts. Hybrid models like SegFormer \cite{a:32} and Uformer \cite{inp:20} address this by integrating convolutional layers with transformers to handle local and global features. Although vision transformers address CNN's limitations in capturing long-range pixel dependencies, they require large datasets, and self-attention limits their ability to learn local pixel relations \cite{a:49, a:48}.

Decoders are crucial in segmentation, transforming low-resolution features into detailed pixel-level predictions. The U-Net, with its skip connections, improves spatial information retention, but early decoders relying on simple upsampling \cite{a:7, inp:12} often generated coarse segmentation maps. Attention based decoders \cite{a:6, a:14, a:24} further enhanced multi-scale feature fusion, but they struggled with precise boundary delineation.

Although these advancements have brought improvements, there remains a significant need for methods that can better capture pixel dependencies across all dimensions and produce accurate segmentation. For motivation, encoders have limitations, and beyond a point, further improvements do not enhance accuracy. Experiments in this study show that even high-end pre-trained encoders may not achieve SOTA accuracy. Thus, we prioritized optimizing the decoder, testing it with both pre-trained and non-pretrained encoders. In this study, we focus on segmenting biomarkers such as nuclei and mitochondria in medical images, which pose significant challenges due to their size variation, unclear boundaries, and complex appearances. Our primary contribution is a significant increase in segmentation accuracy, achieved through a novel decoder design that accurately segments diverse biomarkers while overcoming the limitations of existing approaches. Our main contributions are as follows:

\begin{itemize}

\item We have developed a novel \textbf{Multiscale Convolution Attention with Depth-to-Space (MCADS)} decoder for medical image segmentation. The MCADS decoder leverages the multi-stage feature representation of vision encoders while learning multiresolution spatial representations. It incorporates novel attention modules to suppress unnecessary information. Additionally, it utilizes a \textbf{Depth-to-Space Upsampling Block (DSUB)} in the distal decoder layers to prevent information loss during upsampling.

\item We proposed new attention mechanisms, including the \textbf{Residual Linear Attention Block (RLAB)}, which effectively integrates and refines features between the skip connections of the encoder and the feature maps of the decoder to prioritize the most relevant information. Additionally, the \textbf{Channel and Spatial Attention Block (CASAB)} was introduced to emphasize important channels and focus on relevant spatial regions.

\item We proposed modifications for the U2-Net encoder, effectively capturing rich multi-scale local and global contextual information to complement our decoder for the semantic segmentation task. The encoder outperforms SOTA encoders, but it is especially strong when combined with MCADS, obtaining 2-3\% higher accuracy than SOTA.

\item We conducted extensive experiments, and our evaluation demonstrates that these advancements collectively improve the performance of our proposed method compared to SOTA models.
\end{itemize}

\section{Related Work}
\textbf{CNN's} have been crucial as encoders for managing spatial relationships in images. In segmentation, U-shaped networks are favored for their efficient encoder-decoder structure, with U-Net pioneering the use of skip connections to integrate multi-level information in decoding. Hover-Net \cite{a:47} further enhances this by using the distance of nuclear pixels from their centers of mass to capture detailed information. Studies \cite{inp:10, inp:11, inp:12, inp:17} have demonstrated that incorporating features from multiple deep layers can improve results. Variants of UNet like U2-Net \cite{a:5} incorporate multi-scale feature fusion to capture local and global context; UNet++ \cite{inp:15}, UNet 3+ \cite{inp:19} utilize dense and full-scale skip connections enhancing feature map integration, and incorporate deep supervision to further improve decoder performance. Despite improvements, these models still struggle to capture long-range dependencies, sparking interest in attention based models. Examples include MA-Unet \cite{inp:21}, which leverages multi-scale features and attention. Additional attention techniques like spatial and residual attention have been incorporated into other UNet models, such as SCAU-Net \cite{a:33} and RAUNet \cite{inp:22}. The nnU-Net \cite{a:34} model addresses adaptability by enabling self-tuned preprocessing for optimal predictions. SegResNet \cite{inp:segresnet} utilizes a variational autoencoder for training along with a segmentation decoder. Although CNNs are fundamental to many vision applications, their inherent reliance on localized receptive fields constrains their capacity to effectively capture long-range dependencies within images.

\textbf{Transformers} based architectures represent a recent breakthrough in computer vision. For image segmentation, TransUNet \cite{a:21} uses a transformer as an encoder for robust feature representation and a U-Net-like decoder to effectively upsample the features. The Swin-UNet \cite{inp:9} model uses a hierarchical Swin Transformer encoder with shifted windows for context extraction and a symmetric decoder with a patch-expanding layer to efficiently restore spatial resolution during upsampling. SegFormer \cite{a:32} simplifies the decoder with an MLP based approach.
Additionally, UCTrasNet \cite{inp:36} introduces channel-wise cross-fusion and attention mechanisms for skip connections using a transformer backbone. However, these methods rely on large datasets to effectively learn long-range dependencies, and the improvements in segmentation accuracy have been limited. While vision transformers overcome the limitations of CNNs in capturing long-range pixel dependencies, the self-attention used in transformers limits their ability to learn local (contextual) relations among pixels \cite{a:49, a:48}.

Improvement in \textbf{Decoder} design has been explored through various strategies, including multi-scale context aggregation \cite{inp:26}, the integration of dense \cite{inp:15} or full-scale skip connections \cite{inp:19}, deep supervision \cite{a:39, a:40}, efficient spatial reconstruction modules \cite{inp:27, inp:28}, dual decoder designs \cite{inp:16}, and the incorporation of transformer blocks in the decoder \cite{inp:9}. For instance, PolypPVT \cite{a:41} and Yang et al. \cite{a:42} integrate CBAM \cite{inp:29} within their decoding stages. More recently, Transformer CASCADE \cite{inp:30}, MIST \cite{inp:31}, and EMCAD \cite{inp:28} have proposed decoder designs. Following \cite{inp:9}, they integrate features across multiple stages from a transformer encoder by leveraging various attention mechanisms, which are modified and enhanced versions of \cite{inp:1, inp:29, a:6} to achieve notable performance in medical image segmentation. However, several constraints are associated with these approaches, including a focus on maximizing the utility of pre-trained backbones, attention mechanisms that are suboptimal for capturing both short- and long-range dependencies, and upsampling methods that often fall short in fully restoring information. As a result, improvements in segmentation accuracy remain incremental, highlighting the need for more effective solutions.

The \textbf{Segment Anything Model (SAM)} \cite{inp:42} has proven effective in tackling diverse image segmentation challenges. Utilizing sparse and dense prompts, SAM enables robust feature extraction and zero-shot generalization. Recent studies \cite{inp:43, inp:44, inp:39, inp:41, a:50} have explored fine-tuning strategies to adapt its capabilities for biomarker segmentation. However, despite their advanced feature extraction capabilities, refined fine-tuning techniques, and expert prompting strategies, SAM-based approaches demonstrate minimal improvement or fall short of the accuracy achieved by fully trained networks that operate independently of prompting, as evidenced by our evaluation results and insights from recent studies \cite{a:50, a:51, inp:39}. Although SAM-based methods are not the central focus of this study, they are included to provide deeper insight and a more comprehensive perspective.


\section{Method}

\begin{figure*}
\centerline{\includegraphics[width=1\textwidth]{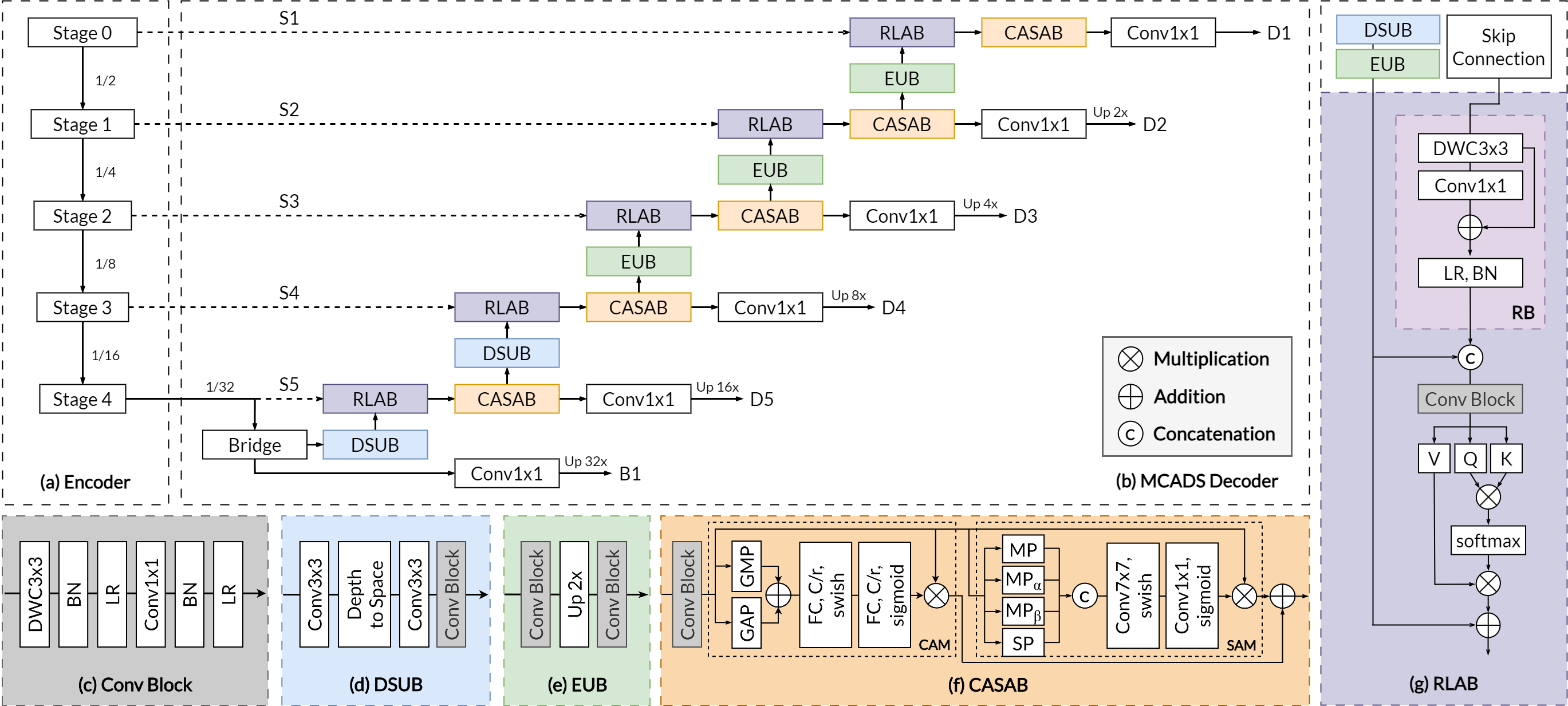}}
\caption{Overview of the proposed architecture: (a) Encoder feature levels at different scales, (b) Multiscale Convolution Attention with Depth-to-Space (MCADS) decoder, (c) Convolution Block, (d) Depth-to-Space Upsampling Block (DSUB), (e) Effective Upsampling Block (EUB), (f) Channel and Spatial Attention Block (CASAB) with Global Max Pooling (GMP), Global Average Pooling (GAP), Max Pooling (MP), Mean Pooling (MP $\alpha$), Min Pooling (MP $\beta$), and Sum Pooling (SP), (g) Residual Linear Attention Block (RLAB) with Residual Block (RB).}
\label{fig:arch}
\end{figure*}

In this section, we will introduce the MCADS decoder that we have proposed. After that, we will delve into detailed explanations of our proposed encoder, the overall architecture, and the loss function.

\subsection{Multiscale Convolution Attention with Depth-to-Space (MCADS) decoder}
The overall decoder architecture is shown in Figure \ref{fig:arch} (b), and the different modules of our decoder are described below.

\subsubsection{Convolution Block}
In our Convolution Block (CB), shown in Figure \ref{fig:arch} (c), we utilize a depthwise 3x3 convolution $DWC_{3x3}$ to extract spatial features, maintaining a lower parameter count than standard convolutions. A subsequent 1x1 convolution $Conv_{1x1}$ adjusts the number of filters. Both convolution layers are followed by Batch Normalization ($BN$) and LeakyReLU activation ($LR$). $LR$ introduces a small slope for negative inputs, preserving a non-zero gradient during inactivity, which helps prevent the dying ReLU problem during training \cite{a:46}. The Convolution Block is formulated as:

\begin{equation} \label{conv_block}
CB = LR(BN(Conv_{1\text{x}1}(LR(BN(DWC_{3\text{x}3}(x)))))
\end{equation}

where $x$ represents the input feature map.

\subsubsection{Depth-to-Space Upsampling Block (DSUB)}
Applying Depth-to-Space $D2S$ early in the decoder preserves detailed information learned by the encoder, which compacts rich features into a smaller spatial space. Whereas EMCAD uses nearest-neighbor upsampling in all stages, which is more prone to information loss. The $D2S$ operation rearranges features into a higher resolution without altering their content, ensuring the decoder can immediately process detailed spatial information. We further validate this design choice through ablation studies, as discussed in Section \ref{sec:abl-dsub}.

In our proposed Depth-to-Space Upsampling Block, shown in Figure \ref{fig:arch} (d), the process begins by determining the number of filters required for upsampling. This is calculated as the product of the output channels and the square of the downsampling factor: $F = C_{\text{prev}} \times 2^d$, where $F$ is the number of filters, $C_{\text{prev}}$ is the number of channels from the previous layer, and $d=2$ is the downsampling factor. This calculation ensures that the feature map is prepared for the subsequent Depth-to-Space ($D2S$) \cite{inp:33} transformation. The $D2S$ operation then rearranges the channels into higher spatial dimensions. A final refinement step is applied. This block is described in Equation~\ref{dsub_block}:

\begin{equation} \label{dsub_block}
\resizebox{0.433\textwidth}{!}{ 
$DSUB = CB(\text{ReLU}(Conv_{3\text{x}3}(D2S(\text{ReLU}(Conv_{3\text{x}3}(x))))))$
}
\end{equation}

\subsubsection{Effective Upsampling Block (EUB)}
The Effective Upsampling Block (EUB), shown in Figure \ref{fig:arch} (e), begins with a convolutional block applied before upsampling. This step reduces noise and irrelevant features learned in earlier stages. Next, the upsampling operation uses bilinear interpolation ($BI$). After upsampling, a second convolutional block is applied to refine the feature map further. This ensures the effective integration of high-level semantic information with the newly increased spatial resolution. Ablation studies in Table \ref{tab:abl-dsub} confirm the computational efficiency and performance of EUB compared to transpose convolutions. The EUB is formulated as:

\begin{equation} \label{eub_block}
EUB = CB(BI(CB(x)))
\end{equation}

\subsubsection{Channel Attention and Spatial Attention Block (CASAB)}
We propose the Channel and Spatial Attention Block (CASAB), shown in Figure \ref{fig:arch} (f), designed to selectively emphasize the most important channels and the most informative features in spatial dimensions. This fusion of channel and spatial attention allows CASAB to capture local spatial information and global channel dependencies, enabling the model to efficiently learn both global and local features. In contrast EMCAD utilize the enhance CBAM \cite{a:38} for spatial and channel attention.
The process of CASAB begins with feature refinement, followed by a Channel Attention Module ($CAM$). In $CAM$, global average pooling ($GAP$) and global max pooling ($GMP$) capture overall feature context and critical activations, respectively. The combination of $GAP$ and $GMP$ provides a more balanced information set, helping the network decide which channels to emphasize or suppress. These pooled features are combined and passed through two $FC$ layers using the $Swish$ activation \cite{a:43} for non-linearity and smooth gradients. The final dense layer with Sigmoid activation generates attention weights that scale the original feature map, emphasizing the most relevant channels. $CAM$ is formulated as:

\begin{equation} \label{cam_module}
\resizebox{0.433\textwidth}{!}{
$CAM(x) = x \, \cdot \, \sigma (FC(Swish(FC(GAP(x) + GMP(x)))))$
}
\end{equation}

where $\sigma$ denotes the Sigmoid activation.

Simultaneously, in the Spatial Attention Module ($SAM$), the input feature map undergoes multiple pooling operations, including mean pooling ($MP$), max pooling ($MP_{\alpha}$), min pooling ($MP_{\beta}$), and sum pooling ($SM$). This captures the presence, prominence, weakest responses, and overall intensity of spatial features, respectively. This is formulated as:

\begin{equation} \label{x_pool}
\resizebox{0.433\textwidth}{!}{
$x_{pool} = Concat(MP(x), MP_{\alpha}(x), MP_{\beta}(x), SM(x))$
}
\end{equation}

These concatenated features are refined using a 7x7 depthwise convolution, capturing broader contextual information. This is followed by $Swish$ and Sigmoid activation, generating a spatial attention map to scale the input feature map and highlight critical spatial regions. The spatial attention module $SAM$ is formulated as:

\begin{equation} \label{sam_module}
\resizebox{0.433\textwidth}{!}{
$SAM(x, x_{pool}) = x \, \cdot \, \sigma (Conv_{1\text{x}1}(Swish(Conv_{7\text{x}7}(x_{pool}))))$
}
\end{equation}

The outputs of the channel and spatial attention modules are combined using element-wise addition, as represented in:

\begin{equation} \label{sam_cam}
CASAB = CAM(x) + SAM(x, x_{pool})
\end{equation}

\subsubsection{Residual Linear Attention Block (RLAB)}
Our proposed Residual Linear Attention Block (RLAB), as shown in Figure \ref{fig:arch} (g), is designed to effectively integrate and refine features between the skip connections of the encoder and the feature maps of the decoder. Furthermore, we validated this design choice through our ablation study, as mentioned in Section \ref{sec:abl-decoder-components}. The RLAB begins by refining the skip connection of the encoder through a specialized Residual Block (RB) as mentioned in Equation \ref{rb_module}. The RB is iterated multiple times, with each iteration involving a residual connection, which ensures that essential information from the encoder is preserved while allowing the network to learn new features. Using residual connections prevents the degradation of feature quality as the input passes through multiple layers. The output of the RB at the $i$-th iteration, $i \in (1, \dots, N)$, where $N$ is the depth of the stage. We configured the number of iterations in the RLAB as [5, 4, 3, 2, 1] for stages 4 through 0, respectively, based on the intuition that the skip connections of deeper layers require more refinement \cite{a:44}. It is formulated as follows:

\begin{equation} \label{rb_module}
\resizebox{0.423\textwidth}{!}{
$x_{RB_i} = BN(LR(x_{RB_{i-1}} + Conv_{1\text{x}1}(Conv_{3\text{x}3}(x_{RB_{i-1}}))))$
}
\end{equation}

After refining the skip connection, the output from the RB is concatenated with the feature map of the decoder. This concatenation merges the high-resolution features of the encoder with the contextually rich, low-resolution features of the decoder, facilitating a comprehensive feature representation that leverages information from both the encoder and decoder pathways, as shown in the following equation:

\begin{equation} \label{concat_enc_dec}
\bar{x} = Concat(x_{RB}, DSUB \hspace{2pt} or \hspace{2pt} EUB)
\end{equation}

Following the concatenation, the combined feature map is refined, after which the feature map is projected into keys, queries, and values using dense layers. The keys and queries compute attention weights through a dot product, then normalized via a softmax function as seen in Equation \ref{attn_calc}. These weights determine how much focus should be placed on different parts of the feature map, allowing the network to prioritize the most relevant information. The resulting weighted values are then added to the original feature map through a residual connection, ensuring that the features of the decoder are retained while being enhanced by the attention mechanism. In contrast, EMCAD employs a modified version of the Attention Gate \cite{a:6}, which filters encoder features by combining them with a gating signal from the decoder to generate attention coefficients. The RLAB block is formulated in Equation \ref{rlab_block}:

\begin{equation} \label{attn_calc}
Attn = softmax\left(\frac{QK^T}{\sqrt{d_{k}}}\right)V
\end{equation}

\begin{equation} \label{rlab_block}
RLAB = \bar{x} + Attn(CB(\bar{x}))
\end{equation}

where $Attn$ is the attention mechanism, and $Q$, $K$, and $V$ are the projection matrices of the query, key, and value, respectively.

This series of operations within RLAB serves as an attention mechanism, which is particularly beneficial for performing precise feature alignment and integration across different levels of the network.

\begin{figure}
\centerline{\includegraphics[width=0.95\columnwidth]{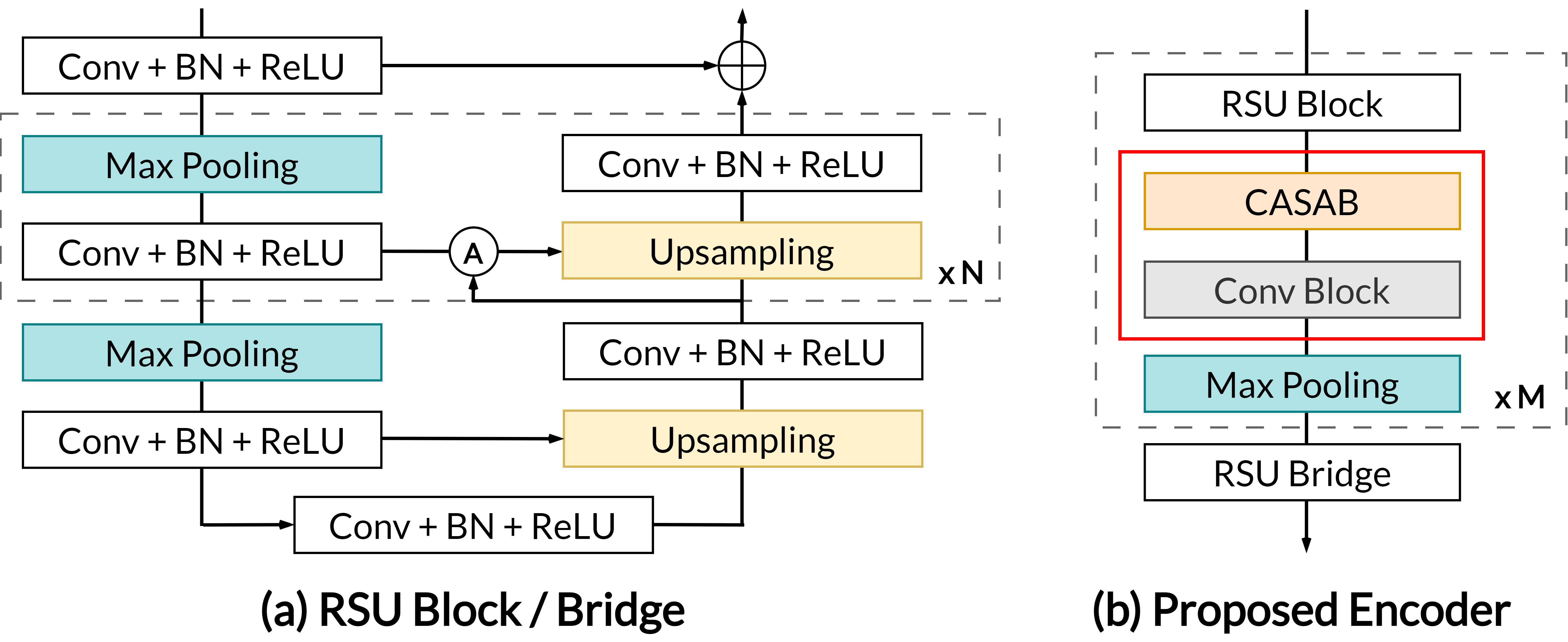}}
\caption{Overall representation of our proposed encoder.}
\label{fig:encoder}
\end{figure}


\subsection{Proposed Encoder}
Following \cite{a:5}, we propose modifications to the residual U-block of U2-Net with our proposed attention mechanisms. We integrated attention mechanisms between the skip connections and the upsampling layers within the residual U-block, enhancing the network's ability to focus on long-range, multi-scale features. This is illustrated in Figure \ref{fig:encoder} (a), where N represents the number of layers in the encoder, equivalent to $L$ in U2-Net. Additionally, after each residual U-block, we implemented CASAB as shown in Figure \ref{fig:encoder} (b), where M repents the encoder stage, which effectively prioritizes the most significant channels and spatial regions, enabling the network to capture fine-grained details while integrating broader contextual information. We validate the effectiveness of our encoder through ablation studies as explained in Section \ref{sec:abl-encoder}.

\subsection{Overall Architecture}
To demonstrate the generalization, effectiveness, and capability of handling multi-scale features in medical image segmentation, we integrate the decoder with the proposed encoder. However, the decoder is highly adaptable and seamlessly compatible with other hierarchical backbone networks, as validated by the ablation studies in Section \ref{sec:abl-encoder}. The decoder processes the features extracted from multiple encoder stages at different scales (S1, S2, S3, S4, S5, and Bridge). It then generates six segmentation maps by applying sigmoid activation to outputs from six stages (B1, D5, D4, D3, D2, and D1), with each map corresponding to a specific encoder stage. Among the prediction maps, D1 is treated as the final output.
We utilize binary cross-entropy loss for training. In the MCADS decoder, the segmentation heads generate six prediction maps, each contributing to a corresponding loss calculation. This approach facilitates deep supervision across all predictions, acting as a form of regularization, strengthening the gradient signal across the network, and guiding each intermediate layer to produce a reasonable segmentation map. This ensures that features useful for segmentation are learned at different levels of abstraction. As a result, the model is better at capturing both local and global context, leading to more accurate segmentation.

The individual losses are summed to form a cumulative loss, which is the primary focus for minimization during the training process. The loss function is represented as follows:

\begin{equation}
\mathcal{L}_{total} = \mathcal{L}_{B_1} + \mathcal{L}_{D_5} + \mathcal{L}_{D_4} + \mathcal{L}_{D_3} + \mathcal{L}_{D_2} + \mathcal{L}_{D_1}
\end{equation}

where $\mathcal{L}$ is the binary cross-entropy loss \cite{inp:37}.
\section{Experiments}

\subsection{Datasets and Evaluation Metrics}
We utilized four publicly available datasets that have also been benchmarked in recent SOTA studies, such as EMCAD (CVPR 2024) \cite{inp:28}, InstaSAM (MICCAI 2024) \cite{inp:39}, FusionU-Net (ACML 2023) \cite{inp:40}, and UCTransNet (AAAI 2022) \cite{inp:36}. The datasets include Multi-organ Nucleus Segmentation (MoNuSeg) \cite{a:1}, the 2018 Data Science Bowl (DSB) \cite{a:3}, Electron Microscopy (EM) \cite{inp:24}, and Triple-negative Breast Cancer (TNBC) \cite{a:2}. These datasets include images captured under different illumination conditions and magnifications, featuring various cell types and modalities. To assess performance, we utilized a range of metrics. For pixel based evaluation, we used Intersection over Union (IoU), Dice coefficient, Precision, Recall, and the False omission rate (FOR). For surface distance based metrics, the 95th percentile of the Hausdorff distance (HD95) and the Average Surface Distance (ASD) were used to provide a comprehensive evaluation.

\subsection{Experimental Setup \& Implementation Details}
To train the model, we utilized the Adam optimizer with a learning rate of $1\text{e}^{-4}$ and a batch size of 8, employing the HeUniform kernel initializer for the convolution operations. The encoder used standard filter sizes commonly employed in encoder-decoder frameworks \cite{a:45}, specifically [64, 128, 256, 512, 512, 512] for stages 0 to 4 and the bridge. We employed end-to-end training without pre-trained networks, using the provided training sets of the MoNuSeg, DSB, and EM datasets. The model was trained entirely from scratch, with over 200 epochs, and best weights were chosen based on performance on the validation set. Both offline and online augmentation were incorporated, generating 256x256 overlapping patches with a stride of 128 for all datasets in the training set. For methods utilizing pre-trained backbones, patches of 224x224 were generated with a stride of 128. To ensure a fair comparison, we reproduced all results for all datasets using their publicly available code and averaged them over five runs. For testing, we created similar overlapping patches, performed predictions, reassembled them, and evaluated the results. The provided test sets of the MoNuSeg and EM datasets were used for evaluation, while for DSB, we followed the evaluation strategy outlined in \cite{inp:28}. Additionally, we provide zero-shot results using the TNBC dataset to offer a comprehensive analysis.


\begin{table}
\centering
\caption{Performance comparison on the MoNuSeg Dataset. Bold indicates the best performance, and underline indicates the second best. We reproduced all results and averaged them over five runs. For optimal performance, we performed fine-tuning on the SOTA methods. Results marked with * were copied from the original papers. '-' indicates unavailable results.}
\label{tab:monuseg}
\resizebox{\columnwidth}{!}{%
\begin{tabular}{llllllll}
\hline
Model & \multicolumn{1}{c}{IoU $\uparrow$} & \multicolumn{1}{c}{Dice $\uparrow$} & \multicolumn{1}{c}{Prec. $\uparrow$} & \multicolumn{1}{c}{Rec. $\uparrow$} & \multicolumn{1}{c}{FOR $\downarrow$} & \multicolumn{1}{c}{HD95 $\downarrow$} & \multicolumn{1}{c}{ASD $\downarrow$} \\ \hline
U-Net \cite{inp:17}       & 59.62       & 73.42       & 74.87       & 73.04          & 0.0766       & 3.6609       & 0.8109       \\
Hover-Net \cite{a:47}     & 62.90       & 77.26       & 77.67       & 77.54          & 0.0689       & 3.8687       & 0.8114       \\
UNet++ \cite{inp:15}      & 69.34       & 81.83       & 75.31       & \textbf{90.29} & 0.0422       & \underline{2.7901} & 0.6458       \\
U2-Net \cite{a:5}         & 68.89       & 81.33       & 77.20       & 86.88          & 0.0444       & 13.445       & 3.0647       \\
nnU-Net \cite{a:34}       & 67.60       & 80.42       & 81.63       & 80.88          & 0.0555       & 3.3357       & 0.6852       \\
FusionU-Net \cite{inp:40} & 66.57       & 79.32       & 73.29       & 87.26          & 0.0417       & 3.9401       & 0.8236       \\
SegResNet \cite{inp:segresnet} & 69.02   & 81.00       & \underline{82.50} & 80.97          & 0.0587       & 13.029       & 2.8829       \\
UCTransNet \cite{inp:36}  & 65.56       & 79.20       & 75.83       & 82.88          & 0.0561       & 18.366       & 4.2000       \\
TransUNet \cite{a:21}     & 67.89       & 80.22       & 81.58       & 79.17          & 0.0610       & 3.5602       & 0.7379       \\
Swin-Unet \cite{inp:9}    & 65.38       & 79.01       & 72.38       & 87.88          & 0.0403       & 3.7871       & 0.7893       \\
PVT-CASCADE \cite{inp:30} & 70.74       & 82.79       & 78.36       & \underline{88.24}    & \underline{0.0362} & 3.0110       & \underline{0.6163} \\
EMCAD \cite{inp:28}       & 71.28 & 82.89 & 82.09 & 83.86      & 0.0488       & 3.1818       & 0.6285       \\
InstaSAM \cite{inp:39}*                  & 59.60        & 78.60        &-              &-                 &-               &-               &-              \\
All-in-SAM \cite{inp:41}*                 & 69.76       & 82.44       &-              &-                 &-               &-               &-              \\
UN-SAM \cite{a:50}*                    & 70.82       & 82.86       &-              &-                 &-               &-               &-              \\
SAC 0-expert \cite{a:51}*                           & \underline{72.61}       & \underline{84.03}       &-              &-                 &-               &-               &-              \\ \hline
Ours                      & \textbf{74.04} & \textbf{85.04} & \textbf{83.78} & 86.43 & \textbf{0.0308} & \textbf{1.8550} & \textbf{0.4810} \\ \hline
\end{tabular}%
}
\end{table}

\begin{table}
\centering
\caption{Performance comparison on the DSB Dataset. We reproduced all results and averaged them over five runs. For optimal performance, we performed fine-tuning on the SOTA methods.}
\label{tab:dsb}
\resizebox{\columnwidth}{!}{%
\begin{tabular}{llllllll}
\hline
Model & \multicolumn{1}{c}{IoU $\uparrow$} & \multicolumn{1}{c}{Dice $\uparrow$} & \multicolumn{1}{c}{Prec. $\uparrow$} & \multicolumn{1}{c}{Rec. $\uparrow$} & \multicolumn{1}{c}{FOR $\downarrow$} & \multicolumn{1}{c}{HD95 $\downarrow$} & \multicolumn{1}{c}{ASD $\downarrow$} \\ \hline
U-Net \cite{inp:17}       & 83.83       & 90.17       & 89.09       & 93.78       & 0.0128       & 7.6797       & 1.9192       \\
Hover-Net \cite{a:47}     & 81.29       & 89.04       & 88.73       & 91.01       & 0.0246       & 7.6289       & 2.1298       \\
UNet++ \cite{inp:15}      & 83.95       & 91.10       & 88.49       & \underline{94.26} & \underline{0.0115} & 10.045 & 2.4359 \\
U2-Net \cite{a:5}         & 82.96       & 90.03       & 90.28       & 91.08       & 0.0163       & 10.869       & 2.3375       \\
nnU-Net \cite{a:34}       & 77.24       & 86.09       & 90.18       & 84.33       & 0.0337       & 9.9094       & 2.7911       \\
FusionU-Net \cite{inp:40} & 80.47       & 88.34       & 93.76       & 85.12       & 0.0332       & 7.6148       & 1.8682       \\
SegResNet \cite{inp:segresnet} & 82.91   & 90.01       & 91.54       & 89.75       & 0.0262       & 7.5580       & 2.0236       \\
UCTransNet \cite{inp:36}  & 85.28       & 91.25       & \underline{93.87} & 90.02       & 0.0158       & 11.514       & 2.5552       \\
TransUNet \cite{a:21}     & 84.72       & 90.90       & 93.37       & 89.86       & 0.0162       & 11.213       & 2.7168       \\
Swin-Unet \cite{inp:9}    & 84.49       & 91.03       & 90.77       & 92.72       & 0.0132       & 8.0415       & 1.9232       \\
PVT-CASCADE \cite{inp:30} & 85.05       & 91.07       & 90.55       & 93.02       & 0.0130       & 7.5427       & 1.7861       \\
EMCAD \cite{inp:28}       & 86.36       & 92.18 & 90.53       & \textbf{94.89} & 0.0544       & \underline{6.4516} & \underline{1.4400} \\
UN-SAM \cite{a:50}*                     & 86.93       & 92.84       & -           & -           & -            & -            & -            \\
SAC 0-expert \cite{a:51}*              & \underline{87.32}       & \underline{93.04}       & -           & -           & -            & -            & -            \\ \hline
Ours                      & \textbf{89.48} & \textbf{94.42} & \textbf{94.94} & 93.94       & \textbf{0.0036} & \textbf{2.6655} & \textbf{0.8559} \\ \hline
\end{tabular}%
}
\end{table}

\begin{table}
\centering
\caption{Performance comparison on the Electron Microscopy Dataset. We reproduced all results and averaged them over five runs. For optimal performance, we performed fine-tuning on the SOTA methods.}
\label{tab:electron}
\resizebox{\columnwidth}{!}{%
\begin{tabular}{llllllll}
\hline
Model &
  \multicolumn{1}{c}{IoU $\uparrow$} &
  \multicolumn{1}{c}{Dice $\uparrow$} &
  \multicolumn{1}{c}{Prec. $\uparrow$} &
  \multicolumn{1}{c}{Rec. $\uparrow$} &
  \multicolumn{1}{c}{FOR $\downarrow$} &
  \multicolumn{1}{c}{HD95 $\downarrow$} &
  \multicolumn{1}{c}{ASD $\downarrow$} \\ \hline
U-Net \cite{inp:17}       & 77.45       & 85.37       & 88.82       & 84.95       & 0.0100       & 46.803       & 10.658       \\
Hover-Net \cite{a:47}   & 79.18       & 88.31       & 91.43       & 85.55       & 0.0084       & 12.624       & 2.1152       \\
UNet++ \cite{inp:15}      & \underline{87.01} & \underline{92.49} & 92.53       & \underline{93.38} & 0.0042       & 4.7633       & 1.1553       \\
U2-Net \cite{a:5}      & 86.31       & 92.01       & 91.92       & 92.88       & \underline{0.0041} & 20.817       & 4.3750       \\
nnU-Net \cite{a:34}      & 85.02       & 91.86       & \underline{94.75} & 89.23       & 0.0064       & 5.2735       & 1.2231       \\
FusionU-Net \cite{inp:40} & 85.92       & 92.37       & 93.99       & 90.87       & 0.0053       & \underline{4.1360} & \underline{1.1485} \\
SegResNet \cite{inp:segresnet}   & 84.51       & 90.87       & 92.79       & 90.13       & 0.0061       & 29.482       & 6.5738       \\
UCTransNet \cite{inp:36}  & 85.99       & 92.32       & 91.55       & 93.24       & 0.0046       & 6.7678       & 1.5115       \\
TransUNet \cite{a:21}   & 84.86       & 91.77       & 94.33       & 89.42       & 0.0062       & 6.1746       & 1.4101       \\
Swin-Unet \cite{inp:9}   & 83.72       & 91.08       & 92.18       & 90.10       & 0.0059       & 9.7707       & 1.8287       \\
PVT-CASCADE \cite{inp:30} & 86.28       & 92.05       & 91.64       & 93.28       & 0.0044       & 26.840       & 5.6181       \\
EMCAD \cite{inp:28}       & 86.02       & 92.43       & 93.97       & 91.02       & 0.0055       & 19.857       & 4.1666       \\ \hline
Ours &
  \textbf{88.89} &
  \textbf{94.09} &
  \textbf{94.81} &
  \textbf{93.43} &
  \textbf{0.0040} &
  \textbf{2.6934} &
  \textbf{0.8552} \\ \hline
\end{tabular}%
}
\end{table}

\begin{table}
\centering
\caption{Zero shot performance comparison on the TNBC Dataset. We reproduced all results.}
\label{tab:tnbc}
\resizebox{\columnwidth}{!}{%
\begin{tabular}{llllllll}
\hline
Model & \multicolumn{1}{c}{IoU $\uparrow$} & \multicolumn{1}{c}{Dice $\uparrow$} & \multicolumn{1}{c}{Prec. $\uparrow$} & \multicolumn{1}{c}{Rec. $\uparrow$} & \multicolumn{1}{c}{FOR $\downarrow$} & \multicolumn{1}{c}{HD95 $\downarrow$} & \multicolumn{1}{c}{ASD $\downarrow$} \\ \hline
U-Net \cite{inp:17}       & 53.41       & 67.55       & \textbf{88.79} & 58.44       & 0.0548       & 21.101       & 5.0472       \\
Hover-Net \cite{a:47}     & 54.89       & 69.20       & \underline{88.68}    & 59.57       & 0.0539       & 23.017       & 5.0716       \\
UNet++ \cite{inp:15}      & 60.94       & 74.53       & 77.89          & 75.20       & 0.0370       & 17.106       & 3.9056       \\
U2-Net \cite{a:5}         & 61.86       & 75.63       & 72.02          & \underline{82.43} & \underline{0.0349} & 14.659       & 3.4034       \\
nnU-Net \cite{a:34}       & 59.35       & 73.06       & 84.57          & 67.63       & 0.0511       & 16.768       & 3.3843       \\
FusionU-Net \cite{inp:40} & 56.80       & 71.44       & 87.84          & 62.21       & 0.0557       & 18.705       & 3.6262       \\
SegResNet \cite{inp:segresnet} & 61.63   & 75.15       & 79.59          & 75.42       & 0.0362       & 15.314       & 3.1723       \\
UCTransNet \cite{inp:36}  & 56.62       & 71.29       & 86.83          & 62.90       & 0.0549       & 18.340       & 3.5854       \\
TransUNet \cite{a:21}     & 59.71       & 73.31       & 85.69          & 66.90       & 0.0518       & 15.273       & 3.2620       \\
Swin-Unet \cite{inp:9}    & 55.87       & 70.20       & 86.78          & 62.13       & 0.0515       & 19.421       & 4.3773       \\
PVT-CASCADE \cite{inp:30} & 62.46       & 76.11       & 82.49          & 74.41       & 0.0379       & \underline{13.506} & \underline{2.7764} \\
EMCAD \cite{inp:28}       & 65.13       & 78.60 & 82.59          & 75.88       & 0.0501       & 15.953       & 3.3419       \\
UN-SAM \cite{a:50}*                     & \textbf{72.27}       & \textbf{83.89}       & -              & -           & -            & -            & -            \\ \hline
Ours                      & \underline{69.16} & \underline{81.48} & 79.63          & \textbf{84.51} & \textbf{0.0276} & \textbf{12.590} & \textbf{1.9611} \\ \hline
\end{tabular}%
}
\end{table}

\subsection{Results}
\begin{figure}
\centerline{\includegraphics[width=0.5\textwidth]{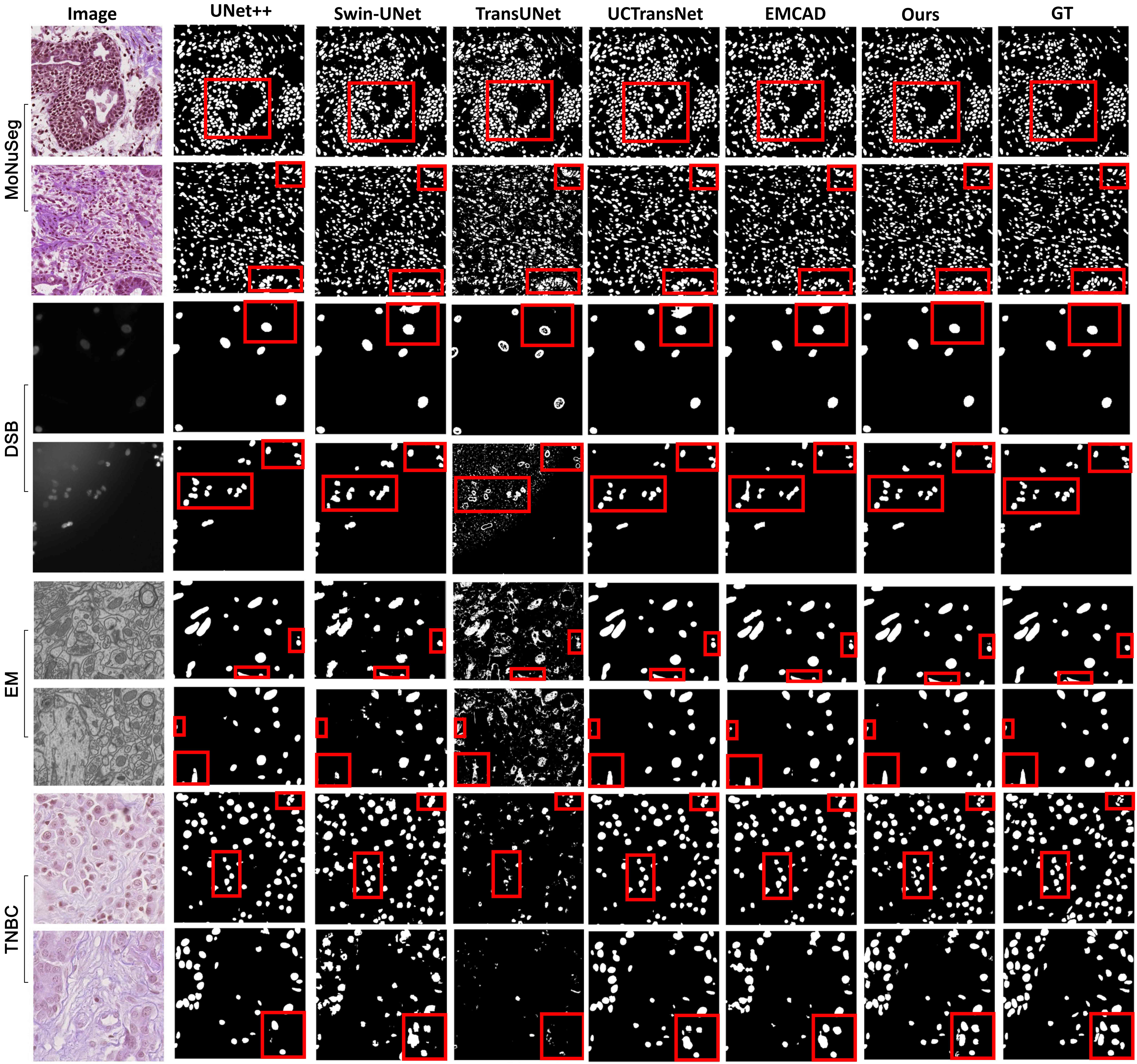}}
\caption{Qualitative Results Comparison: Black pixels represent the background, while white pixels represent the biomarker. The red box highlights regions with false predictions (no prediction or over-segmentation).}
\label{fig:qual}
\end{figure}

We compared our method with SOTA CNN, Transformer and SAM based segmentation approaches focused on biomarker segmentation tasks. Despite not using pre-trained weights like Hover-Net, PVT-CASCADE, and EMCAD, our method outperforms these SOTA models. Our evaluation shows that the proposed method leads in both segmentation accuracy and boundary precision across all datasets, as detailed in Table \ref{tab:monuseg}, \ref{tab:dsb}, \ref{tab:electron}, \ref{tab:tnbc} and in the qualitative results shown in Figure \ref{fig:qual}. We additionally reported the performance of SAM-based methods. Despite their heavy reliance on accurate prompt strategies and fine-tuning, their performance remains moderate and often falls short. Across multiple datasets, methods such as EMCAD and PVT-CASCADE tend to perform moderately but exhibit challenges in boundary precision. UNet++, FusionU-Net and TransUNet offer varying degrees of accuracy, while models like Swin-UNet, UCTransNet, and SegResNet consistently struggle with boundary refinement and overall performance. In zero-shot evaluation, our method exhibits strong robustness. While EMCAD and PVT-CASCADE perform competitively, they lack precision. Models like Swin-UNet, UCTransNet, and SegResNet fall behind, especially in boundary refinement. UN-SAM achieves higher scores due to its fine-tuning on the TNBC dataset, which is not utilized by any other method, including ours. However, the data distribution in the UN-SAM study is unclear. Despite this, we include UN-SAM results to provide an additional evaluation of SAM-based methods. It is worth noting that training on the TNBC dataset could improve our model's accuracy as well.

\begin{figure}
\includegraphics[width=1\linewidth]{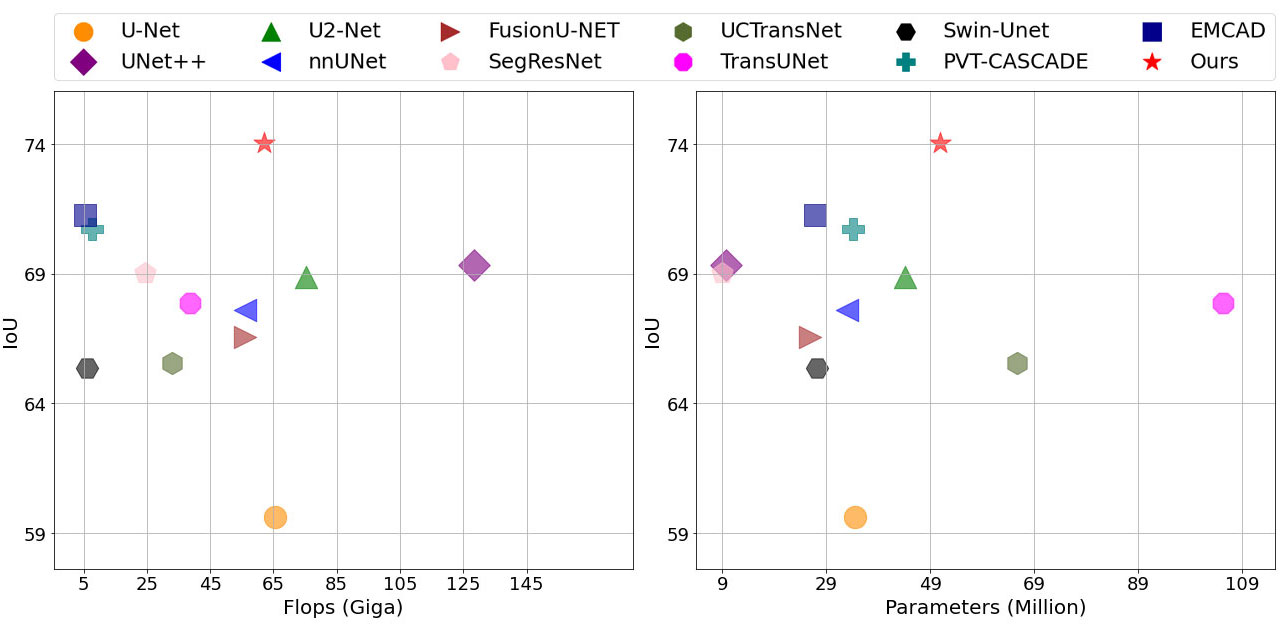}
\caption{Computational Complexity Comparison: IoU vs. GFlops and Parameters.}
\label{fig:complex}
\end{figure}

In terms of \textbf{Computational Cost}, the performance of our method is comparable to other approaches, as shown in Figure \ref{fig:complex}, while delivering a high IoU. This makes our approach ideal for accuracy-critical scenarios. Compared to EMCAD, which has an inference time of 0.1219 seconds, our method takes 0.1398 seconds per sample. The difference of only 0.0179 seconds is negligible, especially since both methods require a GPU for execution, ensuring comparable efficiency in practical scenarios.

\section{Ablation Studies}
\label{sec:ab}
This section discusses the ablation studies we conducted to explore the various aspects of our architecture and experimental framework. For all the ablation studies, we utilize the MoNuSeg dataset train set for training and the test set for evaluation.

\subsection{Effect of proposed modifications on baseline encoder}
\label{sec:abl-encoder}
The ablation study in Table \ref{tab:abl-encoder} evaluates the impact of modifications to the baseline U2-Net encoder, including all other proposed modules. Inner Attention enhances accuracy by focusing on relevant features, while CASAB refines the feature maps by prioritizing important input details. These modifications significantly boost the effectiveness of the encoder, confirming their contribution to the overall performance of the model.

\begin{table}
\centering
\caption{Effect of proposed modifications on baseline encoder}
\label{tab:abl-encoder}
\resizebox{\columnwidth}{!}{%
\begin{tabular}{@{}cclllccc@{}}
\toprule
\textbf{Inner Attention} & \textbf{CASAB} &  &  &  & \textbf{Params} & \textbf{GFlops} & \textbf{IoU} \\ \midrule
\ding{55} & \ding{55} &  &  &  & 50.31 & 61.03 & 72.03 \\
\checkmark & \ding{55} &  &  &  & 50.74 & 61.25 & 72.81 \\
\checkmark & \checkmark &  &  &  & 50.9  & 61.89 & 74.04 \\ \bottomrule
\end{tabular}%
}
\end{table}

\subsection{Effect of DSUB and EUB in MCADS decoder}
\label{sec:abl-dsub}
The ablation study in Table \ref{tab:abl-dsub} evaluates the impact of incorporating DSUB and EUB components into the decoder of the MCADS model. Even when using only Transpose Convolutions with all other modules intact, our method achieves performance comparable to the SOTA methods. However, as EUB and DSUB are gradually introduced, performance steadily improves, particularly after adding DSUB to the first two stages. Beyond this point, the performance gains become minimal while the computational complexity increases significantly. Therefore, we opted to use DSUB only in the first two stages to balance performance with efficiency, allowing further experiments to be conducted within the constraints of our available hardware resources.

\begin{table}
\centering
\caption{Effect of DSUB and EUB in MCADS decoder}
\label{tab:abl-dsub}
\resizebox{\columnwidth}{!}{%
\begin{tabular}{@{}lllllccc@{}}
\toprule
\textbf{Bridge} & \textbf{Stage 4} & \textbf{Stage 3} & \textbf{Stage 2} & \textbf{Stage 1} & \textbf{Param} & \textbf{Flops} & \textbf{IoU} \\ \midrule
ConvTp & ConvTp & ConvTp & ConvTp & ConvTp & 32.36 & 54.59 & 70.52 \\
EUB    & EUB    & EUB    & EUB    & EUB    & 27.45 & 52.51 & 72.23 \\
DSUB   & EUB    & EUB    & EUB    & EUB    & 39.25 & 54.92 & 72.96 \\
DSUB   & DSUB   & EUB    & EUB    & EUB    & 50.9  & 61.89 & 74.04 \\
DSUB   & DSUB   & DSUB   & EUB    & EUB    & 62.99 & 104.4 & 74.10 \\
DSUB   & DSUB   & DSUB   & DSUB   & EUB    & 65.96 & 142.6 & 74.79 \\
DSUB   & DSUB   & DSUB   & DSUB   & DSUB   & 66.73 & 184.9 & 74.88 \\ \bottomrule
\end{tabular}%
}
\end{table}
\subsection{Ablation Study on Encoder-Decoder Configurations}
We compared our proposed decoder MCADS with the baseline decoder EMCAD using two CNN based encoders, namely ResNet-50 and the Modified U2-Net encoder (Ours), as well as two Transformer based encoders, the Multi-axis Vision Transformer (MaxViT-B) and the Pyramid Vision Transformer (PVT v2-B2). These transformer based encoders were selected because they are commonly used in SOTA methods. The ablation study presented in Table \ref{tab:abl-decoder-compare} provides a detailed comparison of various encoder-decoder combinations, highlighting their impact on key performance metrics. 
The results demonstrate that when paired with our decoder, the custom encoder consistently outperforms others, achieving the highest IoU and the lowest HD95 values, indicating superior segmentation accuracy and boundary precision. Notably, even when using different encoder architectures, such as the CNN based ResNet-50 and Transformer based MaxViT-B, our decoder performs better than the EMCAD decoder. This suggests that the architecture of our decoder is robust and effective across various encoder types, further validating the strength of the proposed design.

\begin{table}
\centering
\caption{Comparison of Encoder and Decoder Combinations}
\label{tab:abl-decoder-compare}
\resizebox{\columnwidth}{!}{%
\begin{tabular}{@{}llllllllllcc@{}}
\toprule
\textbf{Encoder} &  & \multicolumn{1}{c}{\textbf{Decoder}} &  &  &  &  &  &  &  & \textbf{IoU} & \textbf{HD95} \\ \midrule
ResNet-50        &  & EMCAD                                &  &  &  &  &  &  &  & 63.36        & 19.916        \\
ResNet-50        &  & Ours                                 &  &  &  &  &  &  &  & 65.59        & 13.072        \\
Ours             &  & EMCAD                                &  &  &  &  &  &  &  & 71.93        & 2.2910        \\
Ours             &  & Ours                                 &  &  &  &  &  &  &  & 74.04        & 1.8550        \\
MaxViT-B         &  & EMCAD                                &  &  &  &  &  &  &  & 70.85        & 2.9933        \\
MaxViT-B         &  & Ours                                 &  &  &  &  &  &  &  & 72.34        & 2.4346        \\
PVT v2-B2        &  & EMCAD                                &  &  &  &  &  &  &  & 71.28        & 3.1818        \\
PVT v2-B2        &  & Ours                                 &  &  &  &  &  &  &  & 72.97        & 2.1755        \\ \bottomrule
\end{tabular}%
}
\end{table}

\subsection{Ablation Study on MCADS Decoder Components}
\label{sec:abl-decoder-components}
We conducted an ablation study to assess the impact of different components in the MCADS decoder, as detailed in Table \ref{tab:abl-components}. The study evaluates the effects of incorporating these components individually and in combination. Initially, without any of these components, the decoder achieves moderate performance. Adding DSUB/EUB significantly improves the IoU, indicating enhanced feature refinement. Introducing RLAB further boosts performance by refining the feature representations through iterative processing. Finally, incorporating CASAB alongside DSUB/EUB and RLAB results in the highest IoU, demonstrating the effectiveness of combining channel and spatial attention mechanisms for optimal segmentation accuracy. This progressive improvement confirms that each component of the MCADS decoder contributes to enhancing the overall performance of the model.

\begin{table}
\centering
\caption{Effect of different components of MCADS decoder.}
\label{tab:abl-components}
\resizebox{\columnwidth}{!}{%
\begin{tabular}{@{}cccccc@{}}
\toprule
\textbf{DSUB / EUB} & \textbf{RLAB} & \textbf{CASAB} & \textbf{Params} & \textbf{GFlops} & \textbf{IoU} \\ \midrule
\ding{55}                 & \ding{55}             & \ding{55}              & 25.45           & 26.59           & 69.03        \\
\checkmark                   & \ding{55}             & \ding{55}              & 43.73           & 36.59           & 70.71        \\
\checkmark                   & \checkmark             & \ding{55}              & 50.71           & 61.79           & 72.17        \\
\checkmark                   & \checkmark             & \checkmark              & 50.9            & 61.89           & 74.04        \\ \bottomrule
\end{tabular}%
}
\end{table}

\section{Conclusion}
We introduces MCADS, a novel decoder designed to reconstruct spatial dimensions and produce accurate segmentation by incorporating RLAB, which integrates multiscale features, and CASAB, which emphasizes important channels and spatial regions. Extensive experiments demonstrate that it consistently outperforms recent SOTA methods across all evaluation metrics. We believe the innovative design will significantly contribute to improving a wide range of medical image segmentation tasks.

\section*{Acknowledgments}
This work was supported by the IITP(Institute of Information \& Coummunications Technology Planning \& Evaluation)-ITRC(Information Technology Research Center) grant funded by the Korea government(Ministry of Science and ICT)(IITP-2025-RS-2023-00259703) and the National Research Foundation of Korea(NRF) grant funded by the Korea government(MSIT)(RS-2025-00573160).


{
    \small
    \bibliographystyle{ieeenat_fullname}
    \bibliography{main}
}


\end{document}